\documentclass[10pt,aps,prl,twocolumn,showpacs,superscriptaddress,nofootinbib,nobibnotes,longbibliography,floatfix]{revtex4-1}

\usepackage[utf8]{inputenc}
\usepackage[T1]{fontenc}
\usepackage{comment}
\usepackage{bm}
\usepackage[normalem]{ulem}\usepackage{mathtools,amsmath,amssymb,amsfonts,mathrsfs,eucal,graphicx,tensor,csquotes,accents,commath,chngcntr,siunitx}
\usepackage[dvipsnames]{xcolor}
\usepackage[unicode]{hyperref}
\hypersetup{colorlinks=true, citecolor=MidnightBlue,
            linkcolor=MidnightBlue, urlcolor=MidnightBlue, linktocpage=true}
\usepackage[normalem]{ulem}

\DeclareMathAlphabet{\mathpzc}{OT1}{pzc}{m}{it}
\definecolor{darkgreen}{rgb}{0.0, 0.6, 0.0}

\begin{document}

\author{Kabir Chakravarti}
\email{chakravarti@fzu.cz}
\affiliation{CEICO, FZU-Institute of Physics of the Czech Academy of Sciences, Na Slovance 1999/2, 182 21
Prague 8, Czech Republic}

\author{Rajes~Ghosh}
\email{rajes.ghosh@iitgn.ac.in}
\affiliation{Indian Institute of Technology, Gandhinagar, Gujarat 382055, India}

\author{Sudipta~Sarkar}
\email{sudiptas@iitgn.ac.in}
\affiliation{Indian Institute of Technology, Gandhinagar, Gujarat 382055, India}

\date{\today}

\title{Formation and Stability of Area Quantized Black Holes }

\begin{abstract}
We investigate the ergoregion instability of area-quantized rotating quantum black holes (QBH) under gravitational perturbation. We show that the instability can be avoided in binary systems that include QBHs if the separation between the inspiralling components at the onset of black hole formation is less than a critical value. We also analyze the formation history of such systems from stellar progenitors and demonstrate that a significant fraction of progenitor masses cannot lead to QBH formation, making it unlikely for LIGO-Virgo black hole binaries to comprise rotating QBHs.
\end{abstract}
\maketitle
\begin{figure}[ht!]
    \includegraphics[width=0.7\columnwidth]{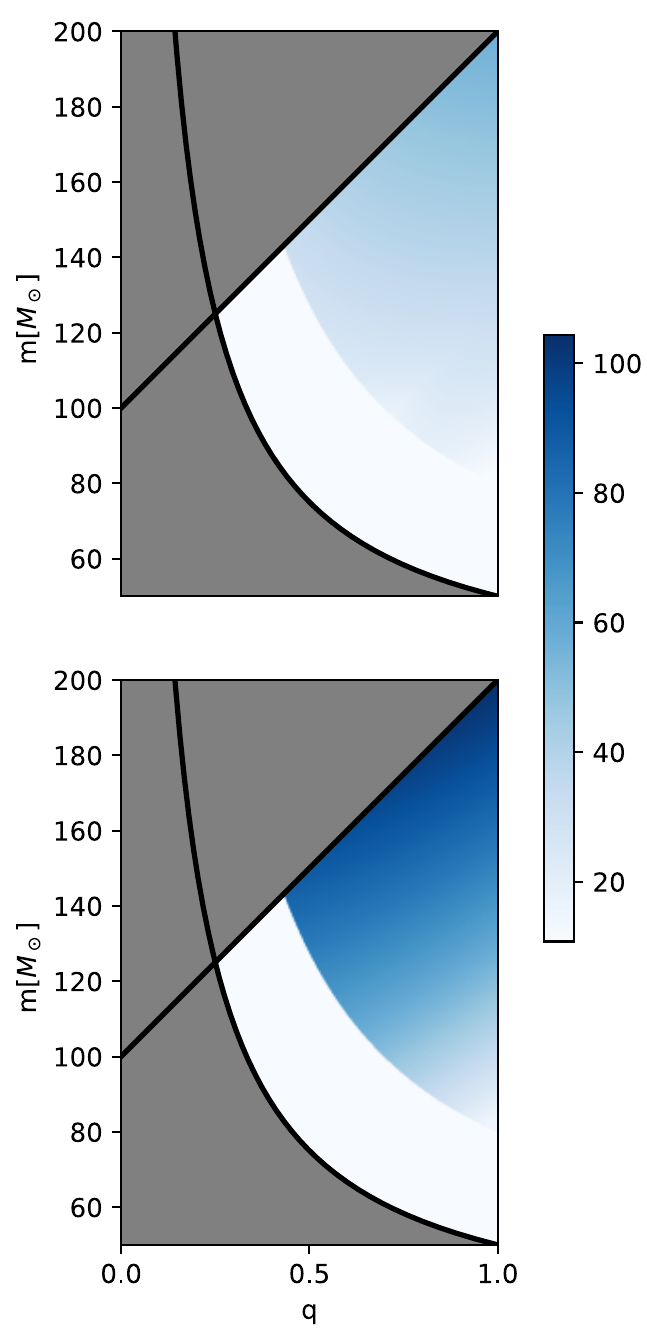}
    \caption{BBH distances $a$ (in units of Sun's radius $R_\odot$, as shown in the side-bar) at the instant of formation as a function of the progenitor configuration. The progenitors start out from a minimum possible distance as explained in the text. Then, they undergo two nova kicks (bottom figure is for the first nova, whereas the top figure is after the second nova) of $100$ km/s as they form QBHs. Note that the grey-shaded portion cannot generate a BBH system.}
    \label{fig:dist_fin_all}
\end{figure}

\noindent{\bf{\em Introduction.}}
Black holes (BHs) are unique laboratory to test our understanding about the fundamental laws of nature. Over time, multiple potential BH candidates have been probed using a variety of observational techniques, including the detection of gravitational waves (GWs) by the LIGO-Virgo collaboration, the observation of BH shadows via the Event Horizon Telescope, and the analysis of other astrophysical phenomena using electromagnetic radiation. These observations have consistently affirmed the presence of massive compact objects that exhibit characteristics akin to those of BHs. Nonetheless, the possibility persists that these entities may in fact be BH mimickers, lacking the defining feature of an event horizon. Hence, a key focus of current astrophysical research is to develop observational methods to distinguish these objects from genuine BHs. \\
Though BHs are solutions to the classical gravitational field equations, their event horizons may reveal interesting features of the yet-to-be-found quantum theory of gravity. One such possibility was proposed by Bekenstein and Mukhanov~\cite{Bekenstein:1974jk, Bekenstein:1995ju}, who considered the idea of a quantum BH (QBH) with horizon area quantized in linear steps (restoring $c$, $G$, and $\hbar$ for the moment),
\begin{equation}
A=\alpha\, \ell_{\rm p}^{2}\, N~.
\end{equation}
Here, $N$ is a positive integer, $\ell_{\rm p}=\sqrt{\hbar\, G/c^{3}}$ is the Planck length, and $\alpha$ depends on the specifics of the quantum gravity. Though there are some heuristic arguments for fixing the value of $\alpha$~\cite{Bekenstein:1974jk, Bekenstein:1995ju, Hod:1998vk, Maggiore:2007nq}, it can also be treated as a phenomenological constant to be measured from observations. Interestingly, besides Bekenstein's original justification based on the adiabatic nature of BH area~\cite{Bekenstein:1974jk, Bekenstein:1995ju}, such discretization might arise as a generic prediction of some proposals of quantum theory of gravity ~\cite{Rovelli:1994ge, Rovelli:1996dv, Ashtekar:1997yu, Ashtekar:2000eq, Agullo:2008yv, Agullo:2010zz, FernandoBarbero:2009ai}. \\
 It was recently recognized that such area-quantized QBHs have distinctive signatures in GW observations because of their selective absorption only at certain characteristic frequencies~\cite{Foit:2016uxn, Agullo:2020hxe}. For a rotating QBH of surface gravity $\kappa$ and horizon angular velocity $\Omega_{\rm h}$, the characteristic frequency associated with the transition $(N, j) \rightarrow (N+n, j+2)$ is given by ~\cite{Agullo:2020hxe},  
\begin{equation}\label{uniform_area}
\omega_{n}=\left(\frac{\alpha\,  \kappa}{8\, \pi}\right)n+2\, \Omega_{\rm h}+\mathcal{O}\big(N^{-1}\big)~.
\end{equation}
Owing to the selective absorption at the horizon, the emitted GWs in the inspiral and post-merger phases of a binary (having at least one QBH as a component) will contain imprints of area quantization. Recent works on tidal heating in the inspiral phase~\cite{Datta:2021row} and echo signals in the ringdown stage~\cite{Cardoso:2019apo} have already shown promising results in this direction. These results suggest that the QBHs may offer an interesting alternative to the standard BH paradigm. Also, if detected, the spectrum of area quantization may provide crucial information about the nature of quantum gravity. All these possibilities have led to a large volume of research aimed at investigating the characteristics of such systems~\cite{Coates:2019bun, Coates:2021dlg, Chakravarti:2021jbv, Chakravarti:2021clm,  Nair:2022xfm}.\\
Despite these exciting advancements, it is imperative to ensure that QBHs do not suffer from any pathology. Otherwise, we can exclude such objects on mere physical grounds. In this work, we study the stability of rotating QBHs under the so-called \textit{ergoregion instability}~\cite{Friedman:1978}, which is linked to the phenomenon of superradiance below a critical perturbing frequency $f_c$~\cite{Brito:2015oca}. Note that a QBH is stable under perturbations with frequencies $f > f_c$, due to the absence of superradiance. However, QBHs behave like perfectly reflecting stars when subjected to perturbations characterized by frequencies lower than $f_c$, rendering instability of the system as shown in Refs.~\cite{Cardoso:2007az, Maggio:2017ivp, Maggio:2018ivz}.\\
Interestingly, for binary systems having at least one QBH component, such instability is avoided if the separation between two inspiralling components at the onset of BH binary is less than a certain critical value corresponding to the critical frequency $f_c$, provided the spin of the QBH formed from the progenitor stars is less than a characteristic value. Using this stability criterion, we find the permissible masses of the progenitor binary stellar systems which can evolve to become binaries with at least one stable QBH. We conclude by showing how stability considerations for QBHs disfavour a significant part of the parameter space for the progenitor masses and provide an upper limit on the mass of BH candidates detected by the LIGO observations to be stable QBHs. \\
\noindent{\bf{\em Ergoregion instability for QBHs.}}
 In the case of rotating QBHs, the incoming perturbation is completely reflected except at the characteristic frequencies $f_n = \omega_n/2\pi$ referred from Eq.~\eqref{uniform_area}. In contrast, at a generic frequency $f \neq f_n$, the surface of the object behaves as a perfectly reflecting boundary with zero transmissivity. Consequently, the reflectivity of a QBH can be modelled as $\mathcal{R}(f_n) = 0$, and away from the characteristic frequencies $\mathcal{R}(f)$ increases smoothly on both sides to reach a value $\mathcal{R}(f_n \pm \Gamma/2) = 1$, where $\Gamma$ denotes the line broadening due to spontaneous Hawking radiation~\cite{Agullo:2020hxe}.\\
Interestingly, if the perturbation frequency $f$ is greater than the lowest transition frequency $f_0 = \omega_0/2\pi$ with $\omega_0=2\Omega_\mathrm{h}$ from Eq.~\eqref{uniform_area}, there will be no ergoregion instability in the absence of superradiance (which requires $\omega < 2\Omega_{\rm h}$). Thus, QBHs are stable for $f > f_0$. However, perturbations below this frequency will lead to ergoregion instability, whose effect will be most prominent in the absence of any surface-absorption~\cite{Maggio:2017ivp}. Then, taking into account the line broadening, a QBH behaves like a perfect reflector below the critical angular frequency $\omega_c=\omega_0-\Gamma/2$. Note, the quantity $\Gamma/2$ denotes the half-width on both sides of a transition line $f_n$ (here, $n=0$). Since the value of $\omega_c$ is independent of $\alpha$, area-quantized BHs suffer from ergoregion instability at perturbing frequencies below $f_c = \omega_c/2\pi$ irrespective of the choice of $\alpha>0$.\\ 
Note that the ergoregion instability is prominently caused by the perturbing GW frequencies in the inspiral phase. These frequencies depend not only on the component QBH's mass and spin, but also on the instantaneous orbital separation. Therefore, unlike the cases presented in Refs.~\cite{Cardoso:2007az, Maggio:2017ivp, Maggio:2018ivz} with quasi-normal modes as the perturbations, we have no bound on the BH's spin to set in the ergoregion instability. In fact, for our case, the stability condition $f > f_c$ can be translated to a bound on the binary orbital separation discussed in the next section. Moreover, critiques may argue that the instability timescale are so large that one may still observe QBH binaries. However, an intuitive argument shows that is not the case. For this purpose, we may follow the analysis of Refs.~\cite{Cardoso:2007az, Maggio:2017ivp, Maggio:2018ivz} and place the near-horizon reflective boundary condition at $r = r_{\rm h} +\delta$, where $r_{\rm h}$ is the location of the Kerr horizon and $\delta \ll r_{\rm h}$. Then, the instability time scale is an order-unity multiple of $r_{\rm h}\, |\log(\delta/r_{\rm h})|$, which is roughly the light-travel time to reach the inner reflecting surface from any finite distance outside the horizon. Thus, for BHs observed by the LIGO with mass not more than $100 M_\odot$ and for reasonable values of $\delta \sim \ell_p$, the instability timescale is always less than a second.\\
Therefore, solely those QBHs can survive the ergoregion instability and manifest as a viable alternative to the classical Kerr BHs, for which the perturbing GW frequency is always above the critical frequency $f_c$. This is only possible if, at the onset of the formation of the BH binary via an astrophysical process, the separation between two inspiralling components (at least one of which is a QBH) is less than a certain critical value corresponding to $f_c$. Thus, the question of stability is then related to the formation history of the BH binary. However, if the formation process leads to non-rotating BHs, the system does not suffer from such instability and hence can not be ruled out on this ground.\\
Note, though we are using gravitational waves as the dominant source of perturbation, there are indeed other possible sources as well (e.g. accreting matter and electromagnetic perturbations etc), which may also add to the ergoregion instability. Moreover, we are only considering a continuous source of perturbation, due to the gravitational radiation, till the ergoregion instability sets in. Nevertheless, once the instability sets in, there is no need for a continuous perturbation to sustain the instability. \footnote{We thank Vitor Cardoso for bringing this point to our notice.}\\
\noindent{\bf{\em Population Analysis.}}\label{sec:pop_an}
Consider a binary system with at least one component being a QBH. Then, there is always a perturbing GW with an angular frequency $2 \Omega$, where $\Omega$ is the average orbital angular frequency. The parameter space of this binary is given by the component masses $m_i$,  spins $\chi_i$, and the binary separation $a$. Here, the index $i$ represents the QBH component(s) in the binary. Since at least one of the binary components is a QBH, we can conclude that any arbitrary configuration of $\{m_i,\chi_i,a\}$ cannot render a stable system if  $\omega < \omega_c$, discussed in the previous section. However, as the average orbital angular frequency $\Omega (t)$ of binaries is a monotonically increasing function of time, the binary will be stable throughout its lifetime if during its formation, 
\begin{equation}\label{BH_stable_cond1}
    2\, \Omega > 2\, \Omega^{(i)}_{\rm h}- \frac{1}{2}\, \Gamma^{(i)}\, ,
\end{equation} 
where $\Gamma$ is the broadening factor of a characteristic absorption line. This condition ensures that after the formation of the QBH, the perturbing frequency is always greater than the critical value. Now, if the RHS of Eq.~\eqref{BH_stable_cond1} is negative, we are guaranteed unconditional stability since the LHS is always positive. This will happen if 
\begin{equation}\label{BH_stable_cond2}
    \frac{2\, \chi^{(i)}}{1 + \sqrt{1-\chi^2_{(i)}}} < m^{(i)}\, \Gamma^{(i)}\, .
\end{equation}
Here, it is understood that the index $i$ refers to the QBH component(s) of the binary. Then, using the fitting function for $m^{(i)}\, \, \Gamma^{(i)}$ used in Ref.~\cite{Agullo:2020hxe}, Eq.~\eqref{BH_stable_cond2} places an upper bound on BH spin required for unconditional stability as $\chi^{(i)} \lesssim 0.0016$.\\
For the remainder of the parameter space, we use Kepler's 3rd law to convert the $\Omega$ inequality in Eq.~\eqref{BH_stable_cond1} into an inequality in $a$. Simultaneously, the restriction to the inspiral phase means $a > 6m$ with $m$ being the total mass of the binary. It is because $a = 6m$ marks the innermost stable circular orbit (ISCO), where the inspiral phase ends to initiate the radial plunge. Combining them together, we get the following inequality,

\begin{equation}\label{a-i_limit}
     6m < a < \Bigg[\frac{m}{\left(\Omega^{\rm G}_{\rm h}\right)^2}\Bigg]^{1/3}\, .
\end{equation}
Here, $\Omega_{\rm h}^{\rm G} = \mathrm{max}\left(2\Omega^{(1,2)}_{\rm h}- \frac{1}{2}\Gamma^{(1,2)}\right)$ for a double QBH binary and $\Omega_{\rm h}^{\rm G} = 2\Omega_{\rm h}- \frac{1}{2}\Gamma$ for a single QBH system. Thus, Eq.~\eqref{a-i_limit} dictates the allowed range of the binary separation such that the inspiralling QBH component(s) is(are) stable.
\begin{figure}[ht!]
    \includegraphics[width=1\columnwidth]{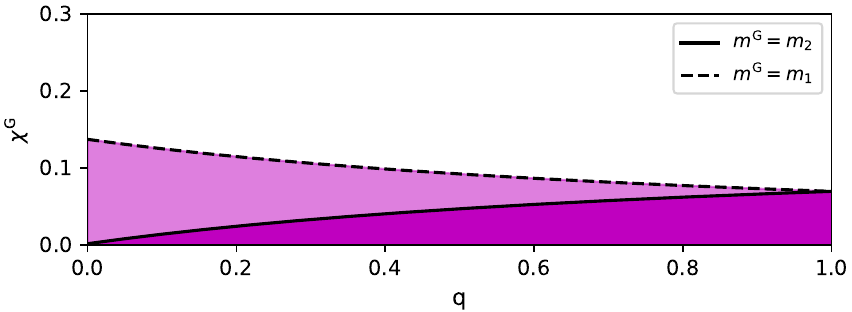}
    \caption{Allowed values of $\chi^{\rm G}$ as a function of mass ratio $q$. The un-shaded region (extends up to the extremal value $\chi^{\rm G} = 1$ of spin) denotes the parameter space where Eq.~\eqref{a-i_limit} fails to hold.}
    \label{fig:spin6m}
\end{figure}
\\Given a fixed the mass ratio $q = m_2/m_1$, Eq.~\eqref{a-i_limit} holds true only for a range of values of spin. The above Fig.~[\ref{fig:spin6m}], plots this threshold spin value ($\chi^{\rm G}$) as a function of $q$. The superscript `G' bears the same meaning as discussed earlier. Thus, a QBH system with $(q, \chi)$ lies in the un-shaded region can not form a stable binary.\\
At this point, it is worth mentioning that both the ISCO radius and Kepler's law receive spin-corrections as the QBH(s) under consideration are Kerr BH(s). However, even for the extremal case, these corrections can at most induce some order-unity modifications and thus, it will not alter the main result (stability/population analysis) of our work. Hence, we shall continue here with Eq.~\eqref{a-i_limit}.
\\Now, we need to know how probable it is for a QBH component to respect the above condition at the onset of the formation of the binary. It is clear that every individual binary configuration would predict a range of stable $a$  that satisfies Eq.~\eqref{a-i_limit}. Adding up those ranges over configurations drawn from a population with characteristic mass and spin distribution, we can generate a probability density plot of $a$. Sophisticated mass distribution functions have been considered in literature \cite{Leyde_2022, Mancarella:2021ecn}, but for simplicity and without loss of generality we consider a uniform and a sharp Gaussian mass distributions as endpoints of a spectrum of distributions. BH spins on the other hand are seeded from a uniform distribution with $0.0016<\chi\leq 1.0$, ensuring no QBH to be unconditionally stable.
\begin{figure}[ht!]
    \includegraphics[width=1\columnwidth]{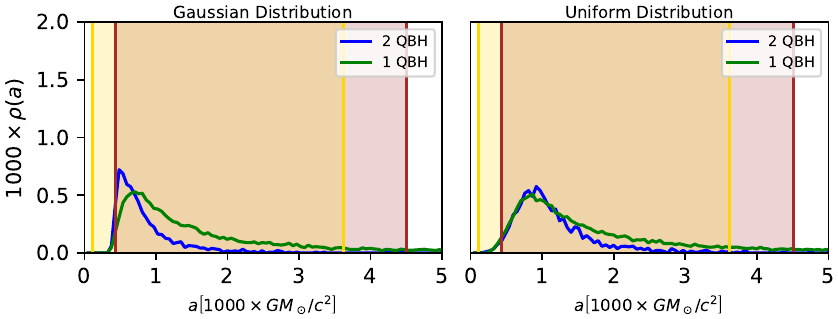}
    \caption{Probability density plot of orbital separation ($a$) necessary for the stability of binary QBHs. Component masses are seeded from Gaussian (top) and uniform distributions. QBH spins are uniformly seeded. The shaded ranges show the calculated $95\%$ CLs of $a$-posteriors of \texttt{GW150914} (brown shade) and \texttt{GW170608} (gold shade).}
    \label{fig:ai}
\end{figure} 
\\The result of such a computation are shown in Fig.~[\ref{fig:ai}] for both single and double QBH systems. As a check of consistency, we have also over-plotted population distribution of $a$ with the posterior of $a_{\rm {max}} = m^{1/3} \left(\Omega^{\rm G}_{\rm h}\right)^{-2/3} $ obtained from \texttt{GW150914} and \texttt{GW170608}. Therefore, we conclude that irrespective of component QBH masses, the formation of a stable binary is possible if the separation $a$ at the onset of the binary formation is in the ballpark of about thousand solar Schwarzschild radii (the peak of the posteriors is near $750$ Km). This is a very small number compared to the average separation between objects trapped in binaries in our local universe, indicating a low probability of stable QBHs in a binary. However, to convert this intuition into result, we now investigate whether there exist progenitor configurations and formation channels which can theoretically give rise to such values of $a$ when the binary BHs (BBHs) are born.\\
\noindent{\bf{\em From progenitor configuration to BBHs.}}\label{sec:proj_to_bbh}
For this work, we \textit{solely} consider potential progenitors that give rise to stellar-mass BBHs typically observed by LIGO. These BHs are thought to be the remnants of core collapsed  massive ($\geq 25 M_\odot$) stars. Based on well-studied models of stellar population synthesis (for example, see Ref.~\cite{Calore:2020bpd}), the most prominent channel of forming a BBH is that progenitor main-sequence stars get trapped in mutual orbit until both its components collapse to BHs, provided the remnants manage to remain in orbit at the endpoint of the entire evolution. Thus, the relevant progenitor configuration space consists of four parameters, namely their masses $m^P_1, m^P_2 (m^P_2\leq m^P_1)$, the binary separation $a^P$, and orbital eccentricity $e^P$. Here, $P$ is an index over the progenitor configuration space. Our goal is to calculate an absolute lower limit of $a^P$ during the orbital evolution via a method of \textit{systematically underestimation} discussed in detail in the \textit{Appendix}. It makes the perturbing GW frequency as large as possible, presenting the greatest possibility of creating a stable QBH binary.\\ 
BHs are not the only remnants possible when progenitor stars die. For the remnants to be just BHs, we impose reasonable cutoffs (see Ref.~\cite{Fryer_2012}) of $m^P_1, m^P_2 \geq 25 M_\odot$ on the progenitor masses. Additionally, hydro-static equilibrium restricts the mass from above. For our purpose, we have taken this to be $100 M_\odot$, i.e., $m^P_1, m^P_2 \leq 100 M_\odot$. The space of progenitors off-limits are indicated by the grey shaded regions of Fig.~[\ref{fig:dist_fin_all}]. We can now pick possible progenitor configurations $P$ and evolve them to obtain the final inter-binary separation, under our method of systematic underestimation. However, we note that as $m^P_1 \geq m^P_2$, the lifespans of the progenitors will not be equal, meaning that in order to get stable double QBHs, the first QBH formed (from $m_1$) would have to be stable as a star-BH system.
\\We evolve our progenitor configurations to see if the stability condition given by Eq.~\eqref{a-i_limit} is obeyed during the star-QBH and QBH-QBH period. It should be noted that an unstable star-QBH system is highly unlikely to evolve to a stable QBH-QBH system, meaning that the instability of a star-QBH system is a much stronger result compared to its QBH-QBH counterpart. However, for making a decisive claim, we have evolved the progenitors to attain the QBH-QBH phase as well. The processes treated under our scheme are namely the binary evolution of the progenitor masses under GW emission, the conservative Roche overflow, and the treatment of one or more successive supernova kicks.\\
As a part of our systematic underestimation scheme, we take the progenitors to start from that separation slightly above the Roche limit which ensures the fastest coalesce rate (because of a greater allowable eccentricity) and no Roche overflow during binary formation. Another important part of our calculation is the effect of two successive supernovas and the associated kicks on our systematic underestimation procedure. A curious reader may follow the \textit{Appendix} for more details.\\  
Finally, we perform our analysis with three values of kicks, namely $50$ (low), $100$ (moderate), and $1000$ km/s (high)~\cite{Hobbs:2005yx}. Here, in Fig.~[\ref{fig:dist_fin_all}], we have only shown the case for the kick $100$ km/s. The plots for other two kick values are presented in the accompanying \textit{Appendix}.\\ 
\noindent{\bf{\em Results and Conclusions.}}\label{sec:res}
Our results of the computations of the binary separations attained after systematic underestimation is presented in 
Fig.~[\ref{fig:dist_fin_all}]. 
First, it is evident that for allowed progenitor configurations, the BBHs are born with $a$ values far outside the $90\%$ CL of the $a$-posterior required to form stable QBHs as suggested by Fig.~[\ref{fig:ai}]. Thus, it can be concluded that progenitors from the allowed regions (shaded blue portion in Fig.~[\ref{fig:dist_fin_all}]) are extremely unlikely to form stable QBH systems, even if every single process in the formation channel was to act favourably. Second, we notice that there are portions of the parameter space in Fig.~[\ref{fig:dist_fin_all}] where systematic underestimation of $a^P(t)$ gives zero. It implies that these configurations may (at least in theory) give rise to the similar separations depicted by Fig.~[\ref{fig:ai}]. Since our calculation is an underestimation, these configurations should be interpreted as the maximum allowable upper limit of the progenitor population that can possibly give rise to stable QBH systems. Among these theoretically possible systems, some of the configurations with parameters $(q, \chi)$ will be ruled out if they happen to lie in the un-shaded region of Fig.~[\ref{fig:spin6m}].\\
Combining all these results together, we calculate the region of progenitor parameter space which can possibly support QBHs (light blue area in Fig.~[\ref{fig:dist_fin_all}]) is about $42.5\%$ of the allowable progenitor parameter space (non-grey area). More interestingly, this ratio is almost independent of the kick values ($\sim 50-1000$ km/s). For example, even a high value of kick like $1000$ km/s can at best make a difference of $\sim 1-2\%$.\\
Finally, we note that as the masses of progenitors capable of generating stable QBHs are restricted and that the remnant masses cannot be larger than those of the progenitors, our results are also indicate an upper bound for the mass of stable QBHs. More quantitatively, we observe from our plots that BBH configurations with total mass $m\geq 120 M_\odot$ and $q\geq 0.6$ are highly unlikely to be QBHs. \\ 
In conclusion, we conducted a detailed, systematic analysis of the possible formation history of area-quantized quantum black holes from the evolution of stellar binary systems. We have arranged the setup so that every aspect of the process of binary evolution conspires to create a stable QBH. Nevertheless, we have found that about $60\%$ of allowed progenitor stellar masses still can not form a stable QBH. In the actual physical situation, it is unlikely that all the physical effects will favor the formation of QBHs. So, we have found only an upper limit of stability; the actual possible range of stellar masses, which can evolve to form a stable QBH, will be much lower than this estimate. Therefore, in conclusion, our work strongly suggests that it is rather unlikely for LIGO-Virgo black hole binaries to comprise of rotating area-quantized QBHs.\\
\begin{acknowledgments}

\noindent
We thank Vitor Cardoso, Jorge Pullin and J.A. de Freitas Pacheco for comments and discussion. The research of K.C is supported by the PPLZ grant (Project number: 10005320/ 0501) of the Czech Academy of Sciences . The research of R.G. is supported by the Prime Minister Research Fellowship (PMRF ID: 1700531), Government of India. S.~S. acknowledges support from the Department of Science and Technology, Government of India under the SERB CRG Grant (CRG/2020/004562). The authors acknowledge the use of the {\it{Noether}} workstation at Indian Institute of Technology Gandhinagar, India. We also thank Sumanta Chakraborty, Parameswaran Ajith, N.V. Krishnendu and Sayak Datta for many useful discussions on various aspects of area quantized black holes.
\end{acknowledgments}
\section*{Appendix}

\noindent{\bf{\em Effect of area-quantization on the line width.}}
Area quantized BHs decay via emission of characteristic frequencies as given by Eq.~(2) of the main text. The available decay channels are thus fewer when compared to their classical counterparts. Moreover, the calculation of the broadening factor $\Gamma \equiv \Gamma_{\rm CBH}$ of the characteristic transition lines (as prescribed in Ref.~\cite{Agullo:2020hxe}) is based on a semi-classical calculation of Hawking radiation by Page~\cite{Page:1976ki}, which for the above reason overestimates the line width and can only be treated as an upper bound on the actual quantum-corrected line width $\Gamma_{\rm QBH}$.
\begin{figure}[ht!]
    \includegraphics[width=1\columnwidth]{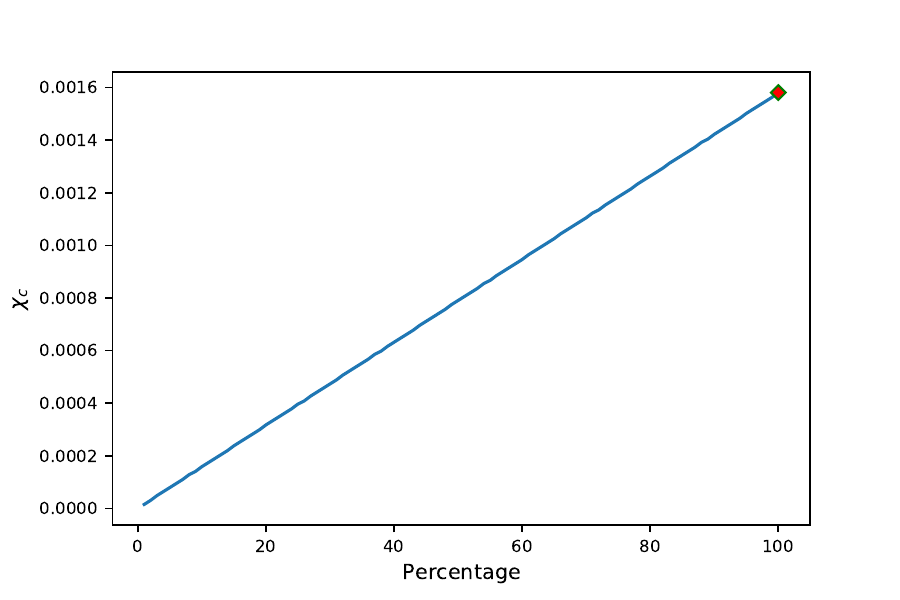}
    \caption{Critical value ($\chi_c$) of spin is plotted as a function of the percentage ($= 100 \times \Gamma_{\rm QBH}/\Gamma_{\rm CBH}$) change of semi-classical line width due to area-quantization. The red diamond on the top-right corner represents the case where $\Gamma_{\rm QBH} = \Gamma_{\rm CBH}$.} 
    \label{fig:chi_crit}
\end{figure} 
\\Since there is no known estimate of the quantity $\Gamma_{\rm QBH}$, one faces an immediate challenge to obtain an upper bound ($\chi \leq \chi_c$) on the QBH spin required for unconditional stability, see Eq.~(4) of the main text discussing the case $\Gamma_{\rm QBH} = \Gamma_{\rm CBH}$. In such a scenario, we may take a simplified assumption that $\Gamma_{\rm QBH}$ is some fraction/percentage of $\Gamma_{\rm CBH}$. In Fig.~[\ref{fig:chi_crit}], we have plotted the critical spin $\chi_c$ as a function of this percentage. It suggests that the value $\chi_c$ always remains small (in fact, bounded above by 0.0016) irrespective of the percentage change. Moreover, we have explicitly checked that this alteration has a negligible effect on our population analysis.\\
\noindent{\bf{\em Initial configuration of progenitors.}}
We highlight briefly our strategy to compute the initial binary progenitor quantities $D^P,\epsilon^P$, given a pair of progenitor masses $m_1^P,m_2^P$. We start with the expression of the Roche radius of the heavier star $m_1^P$ which is approximated to within $1\%$ accuracy by Eggleton's formula \cite{1983ApJ...268..368E}
\begin{align}
r^P_{\rm {RL}}(m_1^P, &m_2^P, D)=   \nonumber \\
D \times &  \left[\frac{0.49 \left(\frac{m_1}{m_2}\right)^{2/3}}{0.60 \left(\frac{m_1}{m_2}\right)^{2/3} + {\rm {log}}\left[1 + \left(\frac{m_1}{m_2}\right)^{2/3}\right]}\right]
\end{align}
for a given inter-binary separation $D$. To get the Roche radius of the smaller star, we just need to swap the labels 2 and 1. We guarantee no-overflow condition at the outset by demanding that the Roche radii $r^P_{k\rm {RL}}$ of each of the components remain larger than their corresponding physical radii $R_k$. It is clearly evident that both inequalities set corresponding lower bounds on $D$. The main sequence mass-radius scaling \cite{1991Ap&SS.181..313D} implies that satisfying the Roche condition at the heavier star automatically ensures it at the smaller star as well. This then, sets for us a minimum distance $D^P_{\rm{RL}}$ between the binary components. However, it is also immediately clear that at such a separation the orbit is forced to be circular if it has to obey the no-overflow criterion. A separation $D \geq D^P_{\rm{RL}}$ can permit the orbit to be eccentric, with an upper limit to the eccentricity $e^P \leq (1 - D^P_{\rm{RL}}/ D)$. Increasing separation reduces rate of quadrupolar emission, while increasing eccentricity increases it.
\begin{figure}[ht!]
    \includegraphics[width=1\columnwidth]{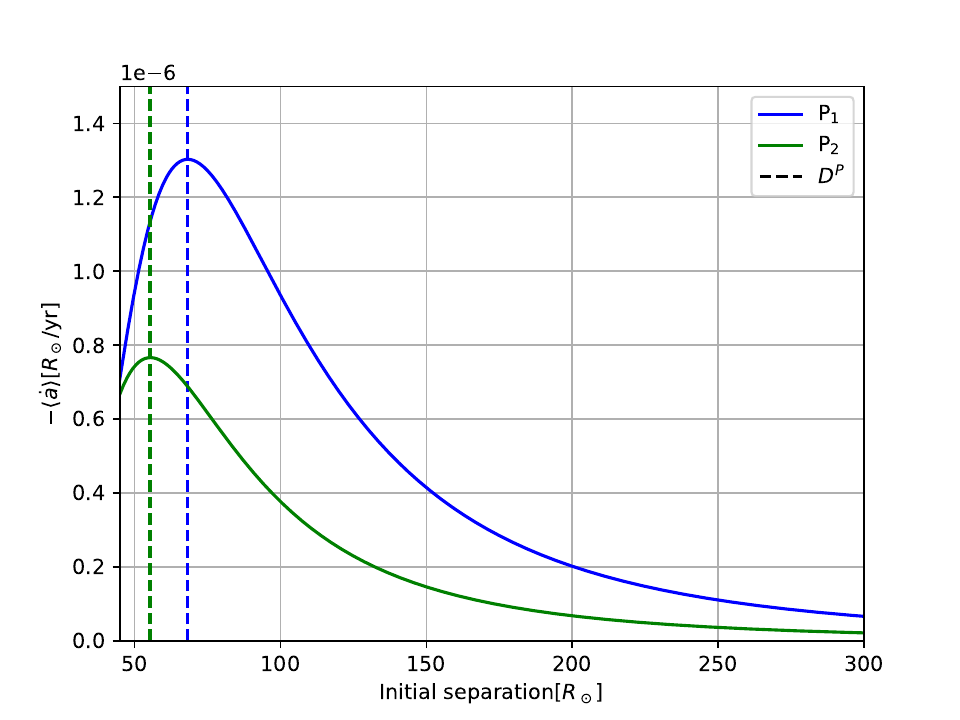}
    \caption{Average coalescence rates $\langle da/dt\rangle$ for two different progenitor mass configurations $P_1 = [100,50] M_\odot$ and $P_2 = [75,35] M_\odot$, as a function of distances above $D^P_{\rm{RL}}$. The maximisation is a result of the antagonistic effects of eccentricity and separation.} 
    \label{fig:coal_rate}
\end{figure} 
It thus turns out that the binary in eccentric orbit with a separation $D^P$ slightly above $D^P_{\rm{RL}}$ is responsible for the fastest coalescence rate, as is demonstrated in Fig.~ [\ref{fig:coal_rate}]. In our algorithm, we compute this separation and eccentricity $\epsilon^p$ corresponding to the maximum average coalescence rate for every pair of progenitor masses, thus giving us $\{m_1^P, m_2^P, D^P, \epsilon^P\}$.   

\begin{figure}[ht!]
    \includegraphics[width=1\columnwidth]{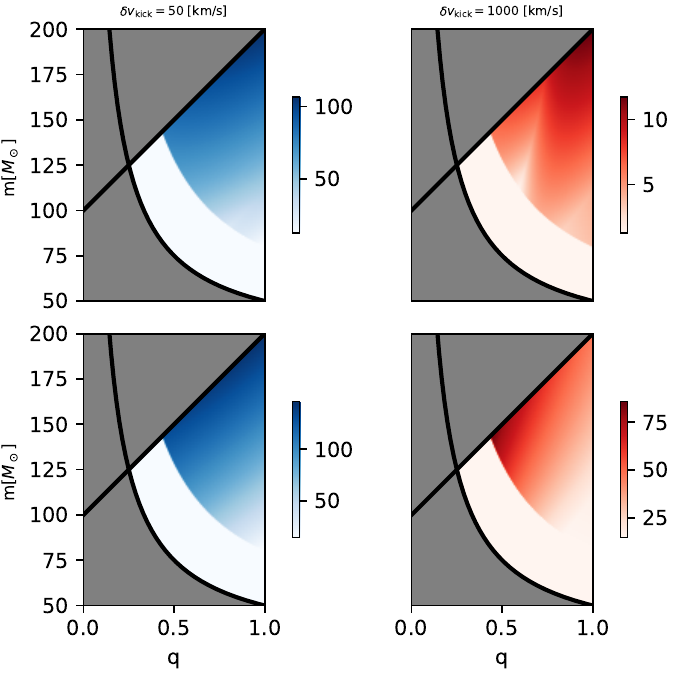}
    \caption{BBH distances $a[R_\odot]$ at birth for low (50 km/s) kicks on the left and high (1000 km/s) kicks. Upper and lower panels denote post-nova 2 and post-nova 1 $a$s  respectively. Note that the high kicks have to be treated non-linearly.}
    \label{fig:dist_fin_all3}
\end{figure}

\noindent{\bf{\em Systematic underestimation.}}
We now highlight our semi-analytical method of systematic underestimation, which allows us to estimate an absolute lower limit of the inter-binary separation  at their endpoint of progenitor evolution for a given initial configuration $\{m_1^P, m_2^P, D^P, \epsilon^P\}$. In the following details, we will suppress the superscript $P$ for brevity.\\
(i) \textit{Incorporating Roche overflow:} Although Roche overflow is assured not to happen initially, it may still occur during the evolution. There are two important factors associated with the overflow that may influence the evolutionary outcome: its nature (conservative or non-conservative), and the associated timescale. Though important for BBH formation~\cite{Belczynski:2017gds}, the non-conservative Roche tidal stripping and the common envelope (CE) evolution phase cannot be accounted for by our simplified method and need numerics which are beyond the scope of this work. However, since the relevance and relative occurrence of these processes are not yet fully understood ~\cite{Inayoshi:2017mrs, Olejak:2021fti}, we can hope to get an indicative (and partial) answer even if we do not take them to account.\\ 
For reasons stated before, we consider the Roche flow to be conservative. Then, consistent with our aim of systematically underestimating $a^P$, the Roche overflow is treated to be instantaneous and is terminated when the composition of the binary becomes symmetric.\\
Let `bR' and `aR' be the labels for configurations before and after the Roche transfer. Then, assuming conservative transfer we end up with 
\begin{equation}\label{roche_loss}
\frac{\delta a}{a}=  \left[\left(\frac{\mu_{\rm{bR}}}{\mu_{\rm{aR}}}\right)^2 - 1\right]\, .
\end{equation}
As the Roche transfers symmetries the configuration, the term in brackets is negative, meaning $\delta a < 0$. This result should now be compared with the loss of separation from quadrupolar GW emission, which (for circular orbits) is given by 
\begin{equation}\label{gw_loss}
\frac{da}{a} = - \frac{64}{5} \frac{\mu m^2}{a^4} dt
\end{equation}
Realistically, both processes can happen simultaneously in nature which require simulations to solve. However, it is immediately apparent that when Eqs.~\eqref{roche_loss} and \eqref{gw_loss} are taken together, the efficiency $(\delta a/a)$ of GW emission to decrease $a$ increases as $-a^{-4}$, while that for the Roche stays constant, for a given progenitor configuration. In a systematic underestimation, one must find the maximally efficient combination of processes that decrease $a$. The corresponding chronological order turns out to be Roche overflow followed by GW emission.\\
(ii) \textit{Incorporating the supernova kicks:} Let us now extend the systematic underestimation to the treatment of the novas and their respective kicks. Supernova simulations demonstrate that following a supernova explosion, the asymmetric ejection of material may impart a resultant natal kick to the supernova remnant. The magnitude and direction of this kick is an intrinsically model dependent quantity, as has been demonstrated by simulations~\cite{Fryer_2012, Bray:2016mab, Mandel:2020qwb}. Additionally, it has also been argued \cite{Belczynski:2016jno, Belczynski:2017gds} that novas seem to disrupt binary progenitor systems and predict rates lower than models which assume CE evolution followed by a direct BH formation without any nova.\\ 
Nonetheless, novas continue to be a relevant phenomenon, given the uncertainties in modelling the event detection rates. A nova kick occurring prograde with the binary orbital motion is likely to increase the inter binary separation and may even disrupt it. Whereas a nova kick retrograde to the orbital motion carries away angular momentum from the system and reduces $a^P(t)$. Continuing with our underestimation procedure, we take each of the novas to be retrograde. The angular momentum carried away by the ejecta is clearly dependent upon mass of the ejecta and remnant as well as on the kick velocity imparted to the remnant. As mentioned before this is intrinsically model dependent. Therefore, for all points in the progenitor space and for each nova therein, we assume a fixed value of the kick velocity. As expected, the range of magnitudes of the imparted kick velocity is speculative as well.\\
Despite the uncertainty, an idea about kick magnitudes can be constructed by the observation of post-nova kick velocity distributions of isolated pulsars which were observed to be fitted by a Maxwellian distribution having standard deviation of $265$ km/s \cite{Hobbs:2005yx}. In our work, we have assumed this velocity distribution to be representative of nova kicks to their remnants. Finally, we perform our analysis with three values of kicks, namely $50$ (low), $100$ (moderate), and $1000$ km/s (high), among which the case for moderate kick has been discussed earlier. Whereas the plots for other two kick values (low and high) are presented in Fig.~[\ref{fig:dist_fin_all3}].\\ 
In addition, note that the nova timescales are much shorter than the orbital timescale of the progenitor binary and hence the nova and its kick are assumed to be instantaneous. This also means that the force on the binary components continue to obey $1/r^2$ law immediately before and just after the nova. We analyse the low to moderate kick regime first. As explained, the nova imparts a kick velocity $\delta v$, while taking away some mass $\delta m$ from the system as ejecta. Remembering $L:= \mu a^2 \Omega$ and Kepler's 3rd Law $\Omega^2a^3 = m$, we have
\begin{equation}\label{low_kick}
    \frac{\delta a}{a} = 2\left(\frac{\delta L}{L}\right) - 2\left(\frac{\delta \mu}{\mu}\right) - \frac{\delta m}{m}
\end{equation}
Notice that for novas $\delta m$ (and therefore $\delta \mu$) is itself negative, so underestimating would mean setting the last two terms to zero, as well as ensuring $\delta a/a <0$ through a maximally retrograde dump $\delta L$ of angular momentum during the kick. It turns out that for both the novas the maximally retrograde $\delta L$ is achieved when $\delta L/L = \delta v / (a\Omega)$. The values of $\delta v$ are then chosen as explained in the main text. Let us now move to the high kick regime, where Eq.~\eqref{low_kick} is modified as
\begin{equation}\label{high_k}
    \frac{\Delta a}{a} = \left(1 + \frac{\Delta L}{\mu a^2\Omega} \right)^2 \left( 1 + \frac{\Delta\mu}{\mu}\right)^{-2} \left( 1 + \frac{\Delta m}{m}\right)^{-1} - 1
\end{equation}
We denote the differential changes now by $\Delta$ to indicate non-linear behaviour. We can again set $\Delta\mu = \Delta m = 0$ initially, as keeping them non-zero  would increase $\Delta a/a$. Once again we need to calculate the maximum retrograde dump $\Delta L$. It is found that this value is equal to $\Delta L/(\mu a^2\Omega) = (1 - m_{\rm {opp}}/ m_{\rm T})$, where $m_{\rm {opp}}$ is the mass of the component opposite to the one having the nova, and $m_{\rm T}$ is the total mass just before the nova. Putting this back into Eq.~\eqref{high_k}, we get
\begin{equation}\label{hk2}
    \frac{\Delta a}{a} = \left(1 - \frac{m_{\rm{opp}}}{m_{\rm T}}\right)^2 - 1
\end{equation}
We note that interestingly from Eq.~\eqref{hk2}, the quantity $\Delta a/a$ now becomes independent of the kick velocity, while underestimation in the non-linear regime. After the first nova the quantity $m_{\rm{opp}}/m_{\rm T} = m_2/m$. Then, after the second nova, this quantity turns out to be $(m_1+\Delta m_1)/(m+\Delta m_1)$, where $\Delta m_1$ is the magnitude of the mass carried away at the first nova. Also, we remind ourselves that in our convention $\Delta m_1 < 0$ A systematic underestimation can further be performed considering that the above fraction increases monotonically with $\Delta m_1$ in the physically viable range $- m_1\leq \Delta m_1\leq 0$. Setting $\Delta m_1 = 0$, we can therefore get the maximum possible of $m_{\rm{opp}}/m_{\rm T} = m_1/m$.



\begin{thebibliography}{100}

\section*{\bf{References}}  

\bibitem{Bekenstein:1974jk}
J. D. Bekenstein, Lett. Nuovo Cim. 11, 467 (1974).


\bibitem{Bekenstein:1995ju}
J. D. Bekenstein and V. F. Mukhanov, Phys. Lett. B 360,
7 (1995), arXiv:gr-qc/9505012.


\bibitem{Hod:1998vk}
S. Hod, Phys. Rev. Lett. 81, 4293 (1998), arXiv:gr-
qc/9812002.


\bibitem{Maggiore:2007nq}
 M. Maggiore, Phys. Rev. Lett. 100, 141301 (2008),
arXiv:0711.3145 [gr-qc].


\bibitem{Rovelli:1994ge}
C. Rovelli and L. Smolin, Nucl. Phys. B 442, 593 (1995),
[Erratum: Nucl.Phys.B 456, 753–754 (1995)], arXiv:gr-
qc/9411005.


\bibitem{Rovelli:1996dv}
C. Rovelli, Phys. Rev. Lett. 77, 3288 (1996), arXiv:gr-
qc/9603063.


\bibitem{Ashtekar:1997yu}
A. Ashtekar, J. Baez, A. Corichi, and K. Krasnov, Phys. Rev. Lett. 80, 904 (1998), arXiv:gr-qc/9710007.


\bibitem{Ashtekar:2000eq}
A. Ashtekar, J. C. Baez, and K. Krasnov, Adv. Theor.
Math. Phys. 4, 1 (2000), arXiv:gr-qc/0005126.
 

\bibitem{Agullo:2008yv}
I. Agullo, J. F. Barbero G., J. Diaz-Polo, E. Fernandez-
Borja, and E. J. S. Villasenor, Phys. Rev. Lett. 100, 211301 (2008), arXiv:0802.4077 [gr-qc].


\bibitem{Agullo:2010zz}
I. Agullo, J. Fernando Barbero, E. F. Borja, J. Diaz-Polo,
and E. J. S. Villasenor, Phys. Rev. D 82, 084029 (2010),
arXiv:1101.3660 [gr-qc].


\bibitem{FernandoBarbero:2009ai}
G. J. Fernando Barbero, J. Lewandowski, and
E. J. S. Villasenor, Phys. Rev. D 80, 044016 (2009),
arXiv:0905.3465 [gr-qc].


\bibitem{Foit:2016uxn}
V. F. Foit and M. Kleban, Class. Quant. Grav. 36, 035006 (2019), arXiv: 1611.07009 [hep-th].


\bibitem{Agullo:2020hxe}
I. Agullo, V. Cardoso, A. del Rio, M. Maggiore,
and J. Pullin, Phys. Rev. Lett. 126, 041302 (2021),
arXiv:2007.13761 [gr-qc].



\bibitem{Datta:2021row}
S. Datta and K. S. Phukon, Phys. Rev. D 104, 124062
(2021), arXiv:2105.11140 [gr-qc].



\bibitem{Cardoso:2019apo}
V. Cardoso, V. F. Foit, and M. Kleban, JCAP 08, 006
(2019), arXiv:1902.10164 [hep-th].


\bibitem{Coates:2019bun}
A. Coates, S. H. Völkel, and K. D. Kokkotas, Phys. Rev.
Lett. 123, 171104 (2019), arXiv:1909.01254 [gr-qc].


\bibitem{Coates:2021dlg}
A. Coates, S. H. Völkel, and K. D. Kokkotas, Class.
Quant. Grav. 39, 045007 (2022), arXiv:2201.03245 [gr-
qc].


\bibitem{Chakravarti:2021jbv}
K. Chakravarti, R. Ghosh, and S. Sarkar, Phys. Rev. D
104, 084049 (2021), arXiv:2108.02444 [gr-qc].



\bibitem{Chakravarti:2021clm}
K. Chakravarti, R. Ghosh, and S. Sarkar, Phys. Rev. D
105, 044046 (2022), arXiv:2112.10109 [gr-qc].



\bibitem{Nair:2022xfm}
S. Nair, S. Chakraborty, and S. Sarkar, Phys. Rev. D
107, 124041 (2023), arXiv:2208.06235 [gr-qc].



\bibitem{Friedman:1978}
J. L. Friedman, Commun. Math. Phys. 63, 243 (1978).


\bibitem{Brito:2015oca}
R. Brito, V. Cardoso, and P. Pani, Lect. Notes Phys.
906, pp.1 (2015), arXiv:1501.06570 [gr-qc].


\bibitem{Cardoso:2007az}
V. Cardoso, P. Pani, M. Cadoni, and M. Cavaglia, Phys. Rev. D 77, 124044 (2008), arXiv:0709.0532 [gr-qc].



\bibitem{Maggio:2017ivp}
E. Maggio, P. Pani, and V. Ferrari, Phys. Rev. D 96,
104047 (2017), arXiv:1703.03696 [gr-qc].



\bibitem{Maggio:2018ivz}
E. Maggio, V. Cardoso, S. R. Dolan, and P. Pani, Phys.
Rev. D 99, 064007 (2019), arXiv:1807.08840 [gr-qc].


\bibitem{Leyde_2022}
K. Leyde, S. Mastrogiovanni, D. Steer, E. Chassande-
Mottin, and C. Karathanasis, Journal of Cosmology and
Astroparticle Physics 2022, 012 (2022).


\bibitem{Mancarella:2021ecn}
M. Mancarella, E. Genoud-Prachex, and M. Maggiore,
Phys. Rev. D 105, 064030 (2022), arXiv:2112.05728 [gr-
qc].


\bibitem{Calore:2020bpd}
F. Calore, A. Cuoco, T. Regimbau, S. Sachdev, and
P. D. Serpico, Phys. Rev. Res. 2, 023314 (2020),
arXiv:2002.02466 [astro-ph.CO].


\bibitem{Fryer_2012}
C. L. Fryer, K. Belczynski, G. Wiktorowicz, M. Dominik,
V. Kalogera, and D. E. Holz, The Astrophysical Journal
749, 91 (2012).


\bibitem{Hobbs:2005yx}
G. Hobbs, D. R. Lorimer, A. G. Lyne, and
M. Kramer, Mon. Not. Roy. Astron. Soc. 360, 974 (2005),
arXiv:astro-ph/0504584.


\bibitem{Page:1976ki}
D. N. Page, Phys. Rev. D 14, 3260 (1976).


\bibitem{1983ApJ...268..368E}
P. P. Eggleton, Astrophys. J. 268, 368 (1983).


\bibitem{1991Ap&SS.181..313D}
O. Demircan and G. Kahraman, Astrophysics and Space
Science 181, 313 (1991).


\bibitem{Belczynski:2017gds}
K. Belczynski et al., Astron. Astrophys. 636, A104
(2020), arXiv:1706.07053 [astro-ph.HE].


\bibitem{Inayoshi:2017mrs}
K. Inayoshi, R. Hirai, T. Kinugawa, and K. Ho-
tokezaka, Mon. Not. Roy. Astron. Soc. 468, 5020 (2017),
arXiv:1701.04823 [astro-ph.HE].


\bibitem{Olejak:2021fti}
A. Olejak, K. Belczynski, and N. Ivanova, Astron. Astro-phys. 651, A100 (2021), arXiv:2102.05649 [astro-ph.HE].




\bibitem{Bray:2016mab}
J. C. Bray and J. J. Eldridge, Mon. Not. Roy. Astron. Soc. 461, 3747 (2016), arXiv:1605.09529 [astro-ph.HE].



\bibitem{Mandel:2020qwb}
I. Mandel and B. Müller, Mon. Not. Roy. Astron. Soc.
499, 3214 (2020), arXiv:2006.08360 [astro-ph.HE].



\bibitem{Belczynski:2016jno}
K. Belczynski et al., Astron. Astrophys. 594, A97 (2016),
arXiv:1607.03116 [astro-ph.HE].



\end{thebibliography}
\end{document}